\documentclass{PoS}

\usepackage{graphicx}                              
\usepackage{amsmath,amsfonts}
\graphicspath{{./pics/}}                       

\usepackage{epsfig}
\usepackage{amssymb}
\usepackage{amsfonts}
\usepackage{bbm}

\newcommand{\preprintline}{\newline
\vskip -4.2cm
\rightline{\parbox{4cm}{\large\rm  HU-EP-08/52\\ DESY 08-158}}
\vspace{3.2cm}}

\title{Higgs mass bounds from a chirally invariant lattice Higgs-Yukawa model with overlap fermions \preprintline}

\ShortTitle{Higgs mass bounds from a chirally invariant Higgs-Yukawa model}

\author{\speaker{Philipp Gerhold}\\
        Institut f\"ur Physik, Humboldt-Universit\"at zu Berlin, 12489 Berlin, Germany\\
        E-mail: \email{gerhold@physik.hu-berlin.de}}

\author{Karl Jansen\\
        DESY, 15738 Zeuthen, Germany\\
        E-mail: \email{Karl.Jansen@desy.de}}

\author{Jim Kallarackal\\
        Institut f\"ur Physik, Humboldt-Universit\"at zu Berlin, 12489 Berlin, Germany\\
        E-mail: \email{Jim.Kallarackal@physik.hu-berlin.de}}

\newcommand{\vs}{\vspace}
\newcommand{\hs}{\hspace}

\newcommand{\bdm}{\begin{displaymath}}
\newcommand{\edm}{\end{displaymath}}
\newcommand{\beq}{\begin{equation}}
\newcommand{\eeq}{\end{equation}}
\newcommand{\bea}{\begin{eqnarray}}
\newcommand{\eea}{\end{eqnarray}}
\newcommand{\bit}{\begin{itemize}}
\newcommand{\eit}{\end{itemize}}
\newcommand{\bc}{\begin{center}}
\newcommand{\ec}{\end{center}}
\newcommand{\re}{\relax{\rm I\kern-.18em R}}
\newcommand{\ID}{\mathbbm{1}}

\newcommand{\fhs}[1]{\mbox{\hs{#1}}}
\newcommand{\ie}{{\it i.e. }}
\newcommand{\Dov}{{\cal D}^{(ov)}}
\newcommand{\D}{\Dov}

\newcommand{\sumFL}{\sum\limits_{i=1}^{N_f}}

\newcommand{\fermiMat}{{\cal M}}

\abstract{We study the parameter dependence of the Higgs mass in a chirally invariant lattice
Higgs-Yukawa model emulating the same Higgs-fermion coupling structure as in the Higgs sector
of the electroweak Standard Model. Eventually, the aim is to establish upper and lower Higgs mass
bounds. Here we present our preliminary results on the lower Higgs mass bound at several 
selected values for the cutoff and give a brief outlook towards the upper
Higgs mass bound.}

\FullConference{The XXVI International Symposium on Lattice Field Theory \\
                 July 14 - 19, 2008\\
                 Williamsburg, Virginia, USA}

\begin{document}

\section{Introduction}
\label{sec:Introduction}

With the existing evidence for the triviality of the Higgs sector of the electroweak Standard Model, 
rendering the removal of the cutoff $\Lambda$ from the theory impossible, physical quantities
in this sector will, in general, depend on the cutoff. Though this restriction strongly 
limits the predictive power of any calculation performed in the Higgs sector, it opens up
the possibility of drawing conclusions on the energy scale $\Lambda$ at which new physics has
to set in, once, for example, the Higgs mass has been determined experimentally.

The main target of lattice studies of the Higgs-Yukawa sector of the electroweak Standard Model 
has therefore been the non-perturbative determination of the cutoff-dependence
of the upper and lower bounds of the Higgs boson mass~\cite{Holland:2003jr,Holland:2004sd} 
as well as its decay properties. There are two main developments which 
warrant the reconsideration of these questions. First, with the advent of the LHC, 
we are to expect that properties of the Standard Model Higgs boson, such as 
the mass and the decay width, will be revealed experimentally. Second, there 
is, in contrast to the situation of earlier investigations of lattice 
Higgs-Yukawa  models~\cite{Smit:1989tz,Shigemitsu:1991tc,Golterman:1990nx,book:Jersak}, 
a consistent formulation of a Higgs-Yukawa model with an exact 
lattice chiral symmetry~\cite{Luscher:1998pq} based on the Ginsparg-Wilson 
relation~\cite{Ginsparg:1981bj}, which allows to emulate the chiral character of the 
Higgs-fermion coupling structure of the Standard Model on the lattice
while lifting the unwanted fermion doublers at the same time.

Before addressing the questions of the Higgs mass bounds and decay properties, we started
with an analytical~\cite{Gerhold:2007yb} and a numerical~\cite{Gerhold:2007gx} investigation of 
the phase structure of the model in order to localize the region in (bare) parameter space 
where eventual calculations of phenomenological interest should be performed. First results
on Higgs mass bounds from chirally invariant lattice Higgs-Yukawa models have already been presented
in \cite{Fodor:2007fn, Gerhold:2007pj}.

In the present paper we study the dependence of the Higgs mass on the model parameters.
We check that the smallest and largest Higgs masses are indeed obtained at vanishing 
quartic Higgs self-coupling and at infinite quartic coupling, respectively, as expected
from perturbation theory. We then present our preliminary results on the cutoff-dependence of the lower Higgs mass 
bound and check the strength of the finite volume effects. Since the aforementioned results
were obtained in the mass degenerate case, \ie with equal top and bottom quark masses,
we also investigate the effect of the top-bottom mass splitting on the Higgs mass, allowing 
ultimately to extrapolate to the - numerically extremely demanding - physical situation, 
where the bottom quark is approximately 40 times lighter than the top quark. We then end 
with a brief outlook towards the upper Higgs mass bound.

\section{The $\mbox{SU}(2)_L\times \mbox{U}(1)_R$ lattice Higgs-Yukawa model}
\label{sec:model}

The model we consider here is a four-dimensional, chirally invariant 
$SU(2)_L \times U(1)_R$ lattice Higgs-Yukawa model~\cite{Luscher:1998pq}, aiming at the 
implementation of the chiral Higgs-fermion coupling structure of the pure Higgs-Yukawa 
sector of the Standard Model reading
\beq
\label{eq:StandardModelYuakwaCouplingStructure}
L_Y = -y_b \left(\bar t, \bar b \right)_L \varphi b_R 
-y_t \left(\bar t, \bar b \right)_L \tilde\varphi t_R  + c.c,
\eeq
with $y_{t,b}$ denoting the top and bottom Yukawa coupling constants.
Here we have restricted ourselves to the consideration of the top-bottom
doublet $(t,b)$ interacting with the complex Higgs doublet $\varphi$ ($\tilde \varphi = i\tau_2\varphi^*,\, \tau_i:\, \mbox{Pauli-matrices}$), 
since the Higgs dynamics is dominated by the coupling to the heaviest fermions. 
For the same reason we also {\it neglect any gauge fields} in this approach. 

The fields considered in this model are one four-component, real Higgs field $\Phi$, being equivalent to the
complex doublet $\varphi$ of the Standard Model, and $N_f$ top-bottom
doublets represented by eight-component spinors $\bar\psi^{(i)}\equiv (\bar t^{(i)}, \bar b^{(i)})$, $i=1,...,N_f$.

The chiral character of the targeted coupling structure~(\ref{eq:StandardModelYuakwaCouplingStructure}) 
is preserved on the lattice by constructing the fermionic action $S_F$ from the Neuberger overlap 
operator~\cite{Neuberger:1998wv} according to
\bea
S_F &=& \sumFL\,
\bar\psi^{(i)}\, \fermiMat\, \psi^{(i)}, \\
\label{eq:DefYukawaCouplingTerm}
\fermiMat &=& \D + 
P_+ \phi^\dagger \fhs{1mm}\mbox{diag}\left(\hat y_t,\hat y_b\right) \hat P_+
+ P_- \fhs{1mm}\mbox{diag}\left(\hat y_t,\hat y_b\right) \phi \hat P_-,
\eea
where the Higgs field $\Phi_n$ was rewritten as a quaternionic, $2 \times 2$ matrix 
$\phi_n = \Phi_n^0\ID -i\Phi_n^j\tau_j$, with $n$ denoting the site index of the $L_s^3\times L_t$-lattice and $\vec\tau$ 
the vector of Pauli matrices acting on the flavour index of the fermionic doublets.
The left- and right-handed projection operators $P_{\pm}$ and the modified projectors $\hat P_{\pm}$
are given as
\bea
P_\pm = \frac{1 \pm \gamma_5}{2}, \quad &
\hat P_\pm = \frac{1 \pm \hat \gamma_5}{2}, \quad &
\hat\gamma_5 = \gamma_5 \left(\ID - \frac{1}{\rho} \D \right),
\eea
with $\rho$ being the radius of the circle of eigenvalues in the complex plane of the free
Neuberger overlap operator~\cite{Neuberger:1998wv}.
This action now obeys an exact $\mbox{SU}(2)_L\times \mbox{U}(1)_R$ lattice chiral symmetry. For $\Omega_L\in \mbox{SU}(2)$
and $U_R \in U(1)$ the action is invariant under the transformation
\bea
\label{eq:ChiralSymmetryTrafo1}
\psi\rightarrow  U_R \hat P_+ \psi + \Omega_L \hat P_- \psi,
&\quad&
\bar\psi\rightarrow  \bar\psi P_+ \Omega_L^\dagger + \bar\psi P_- U^\dagger_R, \\
\label{eq:ChiralSymmetryTrafo2}
\phi \rightarrow  U_R  \phi \Omega_L^\dagger,
&\quad&
\phi^\dagger \rightarrow \Omega_L \phi^\dagger U_R^\dagger.
\eea
Note that in the mass-degenerate case, \ie $y_t=y_b$, this symmetry is extended to $\mbox{SU}(2)_L\times \mbox{SU}(2)_R$.
In the continuum limit the symmetry (\ref{eq:ChiralSymmetryTrafo1},\ref{eq:ChiralSymmetryTrafo2}) recovers the continuum 
$\mbox{SU}(2)_L\times \mbox{U}_R(1)$ chiral symmetry and the lattice Higgs-Yukawa coupling becomes 
equivalent to (\ref{eq:StandardModelYuakwaCouplingStructure}) when identifying 
\bea
\varphi_n = 
-C\cdot \left(
\begin{array}{*{1}{c}}
\Phi_n^2 + i\Phi_n^1\\
\Phi_n^0-i\Phi_n^3\\
\end{array}
\right),\quad
&
\tilde\varphi_n = i\tau_2\varphi^*_n = 
-C\cdot \left(
\begin{array}{*{1}{c}}
\Phi_n^0 + i\Phi_n^3\\
-\Phi_n^2+i\Phi_n^1\\
\end{array}
\right), \quad
&
\mbox{and} \quad
y_{t,b} = \frac{\hat y_{t,b}}{C} \quad
\eea
for some real, non-zero constant $C$. Note that in absence of gauge fields the Neuberger Dirac operator can be
trivially constructed in momentum space, since its eigenvalues and eigenvectors are explicitly known. This will 
be exploited in the numerical construction of the overlap operator.

Finally, the lattice Higgs action $S_{\Phi}$ is given by the usual lattice $\Phi^4$-action
\beq
\label{eq:LatticePhiAction}
S_\Phi = -\hat\kappa\sum_{n,\mu} \Phi_n^{\dagger} \left[\Phi_{n+\hat\mu} + \Phi_{n-\hat\mu}\right]
+ \sum_{n} \Phi^{\dagger}_n\Phi_n + \hat\lambda \sum_{n} \left(\Phi^{\dagger}_n\Phi_n - 1 \right)^2,
\eeq
which is equivalent to the continuum notation
\bea
\label{eq:ContinuumPhiAction}
S_\varphi &=& \sum_{n} \left\{\frac{1}{2}\left(\nabla^f_\mu\varphi\right)_n^{\dagger} \nabla^f_\mu\varphi_n
+ \frac{1}{2}m^2\varphi_n^{\dagger}\varphi_n + \frac{\lambda}{4!}\left(\varphi_n^{\dagger}\varphi_n\right)^2   \right\},
\eea
with the bare mass $m$ and the bare quartic coupling constant $\lambda$. The connection is
established through a rescaling of the Higgs field and the involved coupling constants according
to
\beq
\label{eq:RelationBetweenHiggsActions}
\varphi_n = -\sqrt{2\hat\kappa}
\left(
\begin{array}{*{1}{c}}
\Phi_n^2 + i\Phi_n^1\\
\Phi_n^0-i\Phi_n^3\\
\end{array}
\right),
\quad
\lambda=\frac{\hat\lambda\cdot 4!}{4\hat\kappa^2}, \quad
m^2 = \frac{1 - 2N_f\hat\lambda-8\hat\kappa}{\hat\kappa}, \quad
y_{t,b} = \frac{\hat y_{t,b}}{\sqrt{2\hat\kappa}}.
\eeq

\section{Numerical results}

The general method we apply to determine the lower and upper Higgs mass bounds
is the numerical evaluation of the whole range of Higgs masses that are producible
within our model in consistency with phenomenology. The latter requirement
restricts the freedom in the choice of the model parameters $\hat \kappa, \hat y_{t,b}, \hat \lambda$ 
due to the phenomenological knowledge of the top and bottom masses, 
\ie $m_{t} \approx 175\, \mbox{GeV}$ and $m_{b} \approx 4.2\, \mbox{GeV}$,
respectively, thus fixing the renormalized Yukawa coupling constants for the top and bottom quark.
Furthermore, the model has to be evaluated in the broken phase, \ie at non-vanishing
vacuum expectation value of the Higgs field, $vev \neq 0$, close to the phase transition 
to the symmetric phase. We use the phenomenologically known
value $vev = 246\, \mbox{GeV}$ to determine the lattice spacing $a$ and thus the physical cutoff $\Lambda$ 
according to
\bea
246\, \mbox{GeV} = \frac{\sqrt{2\hat\kappa}\cdot\langle \hat v \rangle}{\sqrt{Z_G}\cdot a}, \quad&
\Lambda = a^{-1}, \quad&
\hat G_G^{-1}(\hat p^2) = \frac{2\hat\kappa\cdot\hat p^2}{Z_G }
\eea
where $\langle \hat v \rangle$ denotes the bare lattice vev and the Goldstone renormalization constant $Z_G$ 
is obtained from the lattice Goldstone propagator $\hat G_G^{-1}(\hat p^2)$ measured in the simulation with $\hat p^2$ 
denoting the squared lattice momenta. For the numerical evaluation of the model we have implemented a PHMC-algorithm, 
allowing to access the physical situation of odd $N_f$. All results in the following are preliminary and have been obtained at $N_f=1$
with degenerate Yukawa coupling constants, \ie $y_t=y_b$ (unless otherwise stated), tuned to reproduce the phenomenologically known
top quark mass. 
However, we are currently working also on the $N_f=3$ results to account for the colour index (even though
gauge fields are absent here).

\bc
\begin{figure}[htb]
\begin{tabular}{cc}
\includegraphics[width=0.48\textwidth]{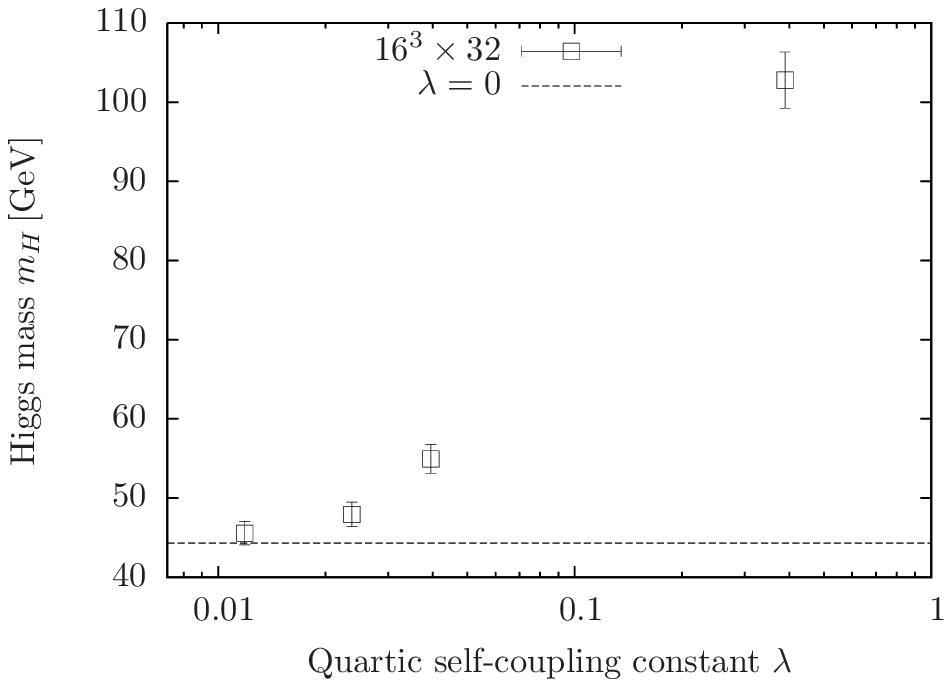}
&
\includegraphics[width=0.48\textwidth]{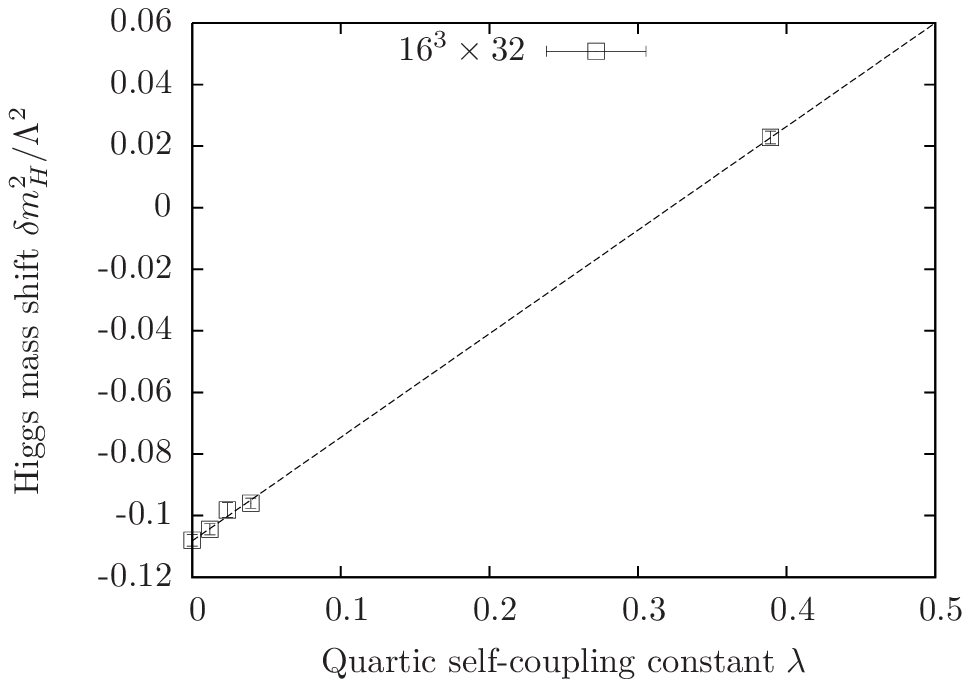}\\
\hs{4mm}(a) & \hs{8mm}(b)  \\
\end{tabular}
\caption{(a) The dependence of the Higgs mass $m_H$ on the quartic self-coupling constant $\lambda$ at 
$\Lambda= 400\, \mbox{GeV}$ on a $16^3\times 32$-lattice for constant Yukawa couplings. 
The dashed line indicates the $\lambda=0$ result.
(b) The corresponding Higgs mass shifts $\delta m_H^2$ versus the quartic coupling constant $\lambda$. 
The dashed line is a linear fit through the points.}
\label{fig:lambdaDependence}
\vs{-2mm}
\end{figure}
\ec

\subsection{$\lambda$-dependence of Higgs mass at $\lambda\ll 1$}
For a given cutoff $\Lambda$ these requirements still leave open an one-dimensional freedom, which can be parametrized in 
terms of the quartic self-coupling constant $\lambda$. However, this remaining freedom can be fixed,
since it is expected from perturbation theory that the lightest
Higgs masses are obtained at vanishing self-coupling $\lambda=0$, and the heaviest masses at infinite coupling $\lambda=\infty$,
according to the one-loop perturbation theory result for the Higgs mass shift~\cite{Veltman:1980mj}
\beq
\label{eq:perturbTheroyResult}
\delta m_H^2 = m_H^2 - m^2 \propto  \left(\lambda - y_t^2 - y_b^2 \right) \cdot \Lambda^2.
\eeq
One should remark here that this argument is not complete, since the phase transition line changes
for varying $\lambda$, thus making the bare Higgs mass $m$ a function of $\lambda$ for fixed cutoff $\Lambda$
and constant Yukawa coupling.
In fact, the bare mass decreases with increasing self-coupling~\cite{Gerhold:2007yb} (in the weak coupling regime),
contributing to the $\lambda$-dependence of $m_H$ with opposite sign as compared to (\ref{eq:perturbTheroyResult}). 
In Fig.~\ref{fig:lambdaDependence}a we therefore check that the lightest Higgs masses are - despite the latter effect - nevertheless 
obtained at vanishing self-coupling, thus allowing to restrict the search for the lower Higgs mass bounds to the 
setting $\lambda = 0$ in the following. The expected linear behaviour in the mass shift with increasing $\lambda$ is 
clearly observed in Fig.~\ref{fig:lambdaDependence}b.

\bc
\begin{figure}[htb]
\begin{tabular}{cc}
\includegraphics[width=0.48\textwidth]{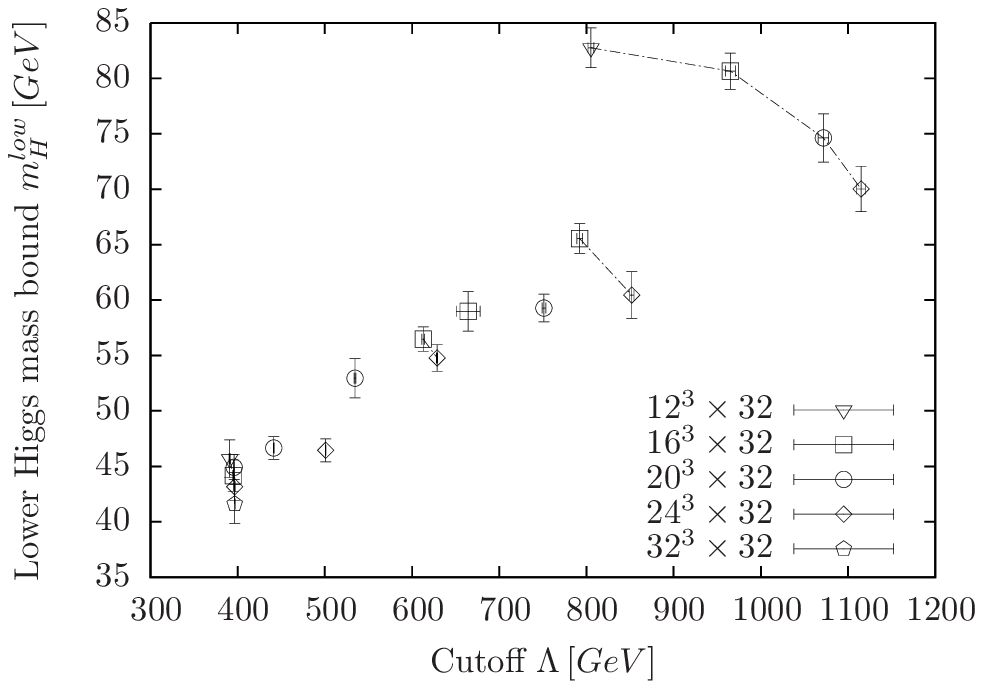}
&
\includegraphics[width=0.48\textwidth]{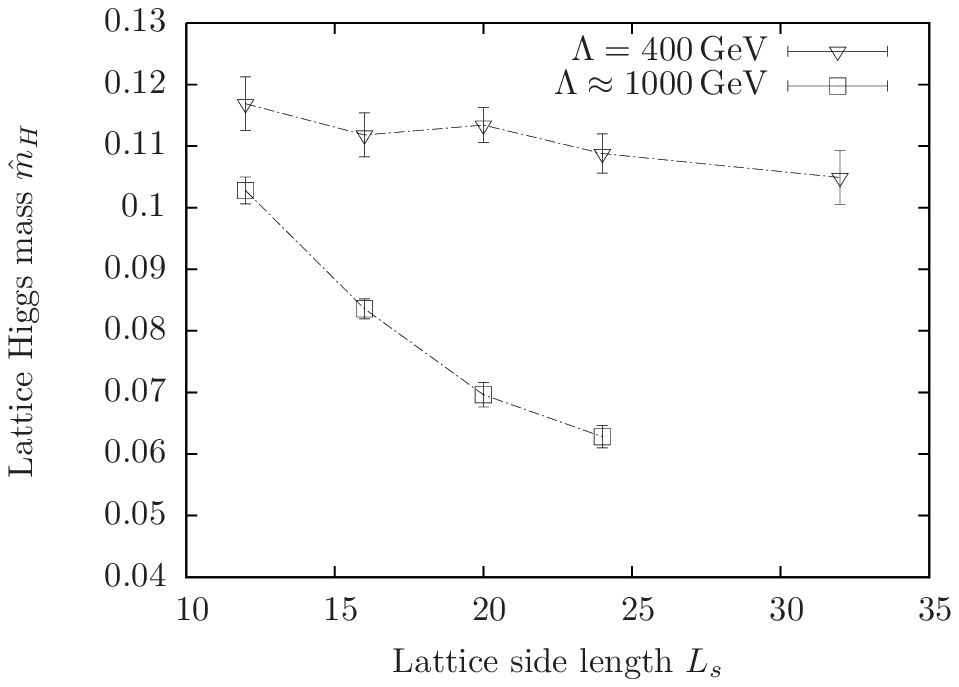}\\
\hs{4mm}(a) & \hs{8mm}(b)  \\
\end{tabular}
\caption{(a) The lower Higgs mass bound $m_H^{low}$ versus the cutoff $\Lambda$ determined on several lattice sizes.
To illustrate finite volume effects, simulations have been rerun with identical parameter sets but different
lattice sizes. Runs with same parameter sets are connected via dashed lines to guide the eye.
(b) Dependence of lattice Higgs masses $\hat m_H$ on the lattice size $L_s$ at $\Lambda= 400\, \mbox{GeV}$ and
$\Lambda\approx1000\, \mbox{GeV}$.}
\label{fig:LowerBound}
\vs{-2mm}
\end{figure}
\ec

\subsection{Cutoff-dependence of lower Higgs mass bound and finite volume effects}

For the determination of the cutoff-dependence of the lower Higgs mass bound we evaluate
the Higgs mass at $\lambda=0$ for several values of $\Lambda$. Two restrictions limit the
range of accessible energy scales: on the one side all particle masses have to be small 
compared to $\Lambda$ to avoid cutoff-effects, on the other side all masses have to be large
compared to the inverse lattice size to avoid finite volume effects. As a minimal requirement 
we demand here that all particle masses $\hat m$ in lattice units fulfill $\hat m < 0.5$ and $\hat m\cdot L_{s,t}>2$. 
For a lattice with side lengths $L_s=L_t=32$, a degenerate top/bottom quark mass of $175\, \mbox{GeV}$, 
and Higgs masses ranging from $40$ to $70$ GeV one can access energy scales $\Lambda$ from $350\,\mbox{GeV}$ 
to approximately $1100$ GeV. In Fig.~\ref{fig:LowerBound}a
we show the obtained Higgs masses versus the cutoff $\Lambda$. To illustrate the influence of the 
finite lattice volume we have rerun some of the simulations with exactly the
same parameter settings but different lattice sizes. Those results belonging to the same parameter
sets are connected by lines to guide the eye.
While the finite volume effects are mild at $\Lambda=400\, \mbox{GeV}$ with $\hat m_H\cdot L_{s,t}>3.2$ on the 
$32^4$-lattice, the vev, and thus the associated cutoff $\Lambda$, as well as the Higgs mass itself vary strongly with 
increasing lattice size $L_s$ at $\Lambda\approx 1000\,\mbox{GeV}$ as can be seen in Fig.~\ref{fig:LowerBound}b. 
Larger lattices are required here to determine the Higgs mass reliably also at this energy scale.

\subsection{Dependence of lower Higgs mass bound on top-bottom mass-splitting}

\begin{figure}[htb]
\begin{tabular}{cc}
\includegraphics[width=0.48\textwidth]{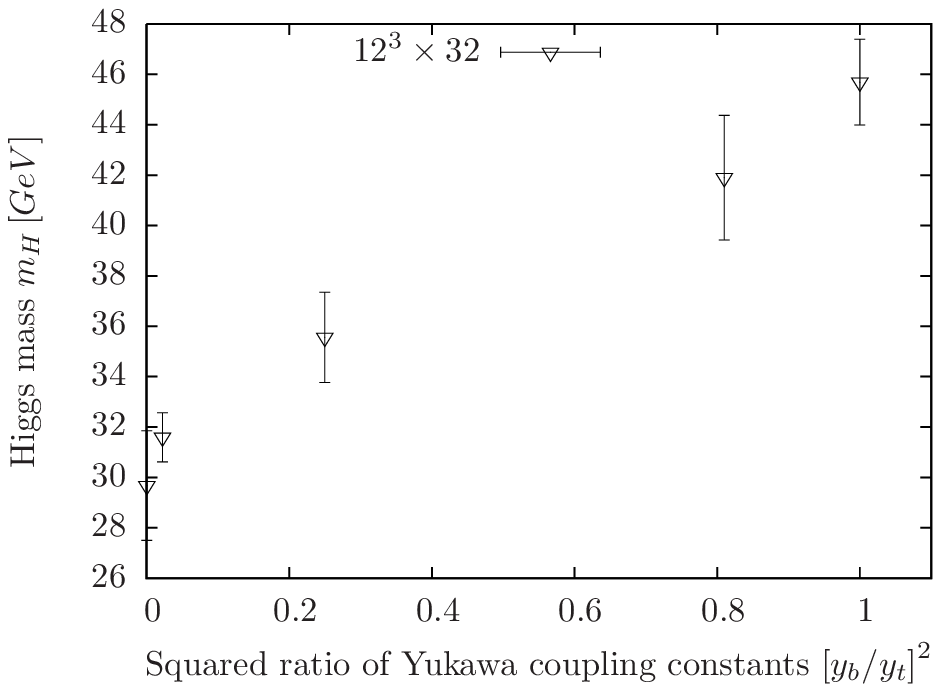}
&
\includegraphics[width=0.48\textwidth]{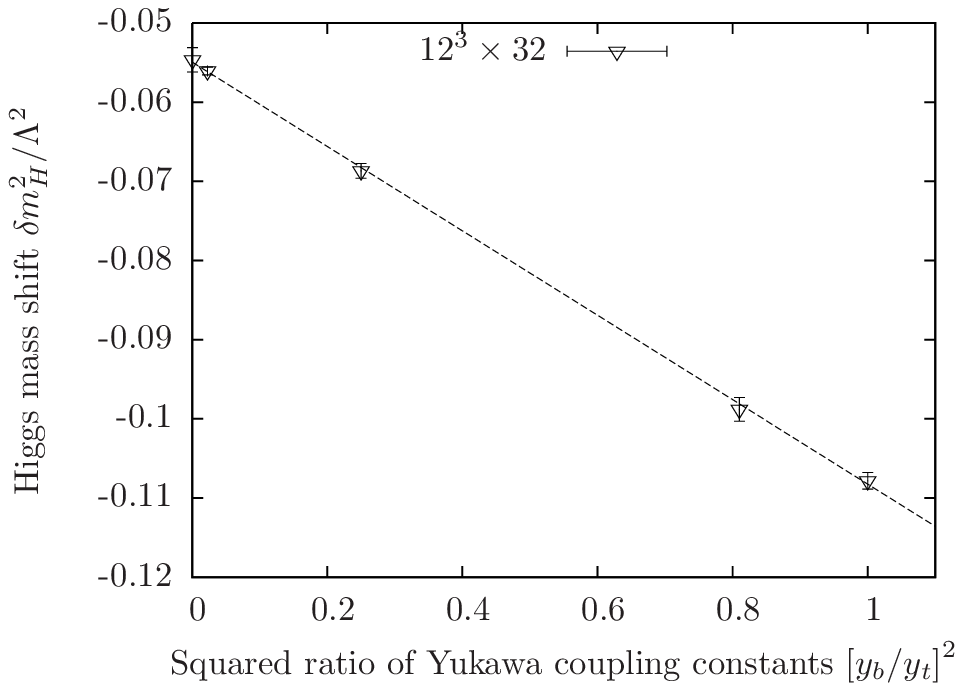}\\
\hs{4mm}(a) & \hs{8mm}(b)  \\
\end{tabular}
\caption{(a) The Higgs mass versus the squared ratio of the top and bottom Yukawa coupling constants 
$\left[y_b/y_t \right]^2$ on a $12^3\times 32$-lattice for constant cutoff $\Lambda=400\,\mbox{GeV}$,
$\lambda=0$, and $m_t=175\,\mbox{GeV}$. 
(b) The corresponding Higgs mass shifts versus $\left[y_b/y_t \right]^2$. The dashed line is
a linear fit through the points.}
\label{fig:SplitDependence}
\vs{-2mm}
\end{figure}

So far, the presented results have been determined in the mass degenerate case, \ie $y_t=y_b$, which is 
easier to access numerically, opening up the question of how the results are influenced when bringing the top-bottom
mass split to its physical value, \ie $m_b/m_t\approx 0.024$. From (\ref{eq:perturbTheroyResult}) one
expects the Higgs mass shift $\delta m_H^2$ to grow quadratically with decreasing $y_b$ and that is
exactly what is observed in Fig.~\ref{fig:SplitDependence}b. Here the top quark mass, the quartic coupling, and 
the cutoff are held constant, while lowering $m_b$ to its physical value.
However, the Higgs mass itself does not increase but {\it decrease} with decreasing $y_b$ as shown in 
Fig.~\ref{fig:SplitDependence}a. This is because the first effect in the mass shift is over-compensated 
by the shift in the phase transition line, which is moved towards smaller bare Higgs masses $m$~\cite{Gerhold:2007yb}.

\bc
\ec

\subsection{Outlook towards upper Higgs mass bounds}

\begin{figure}[htb]
\begin{tabular}{cc}
\includegraphics[width=0.48\textwidth]{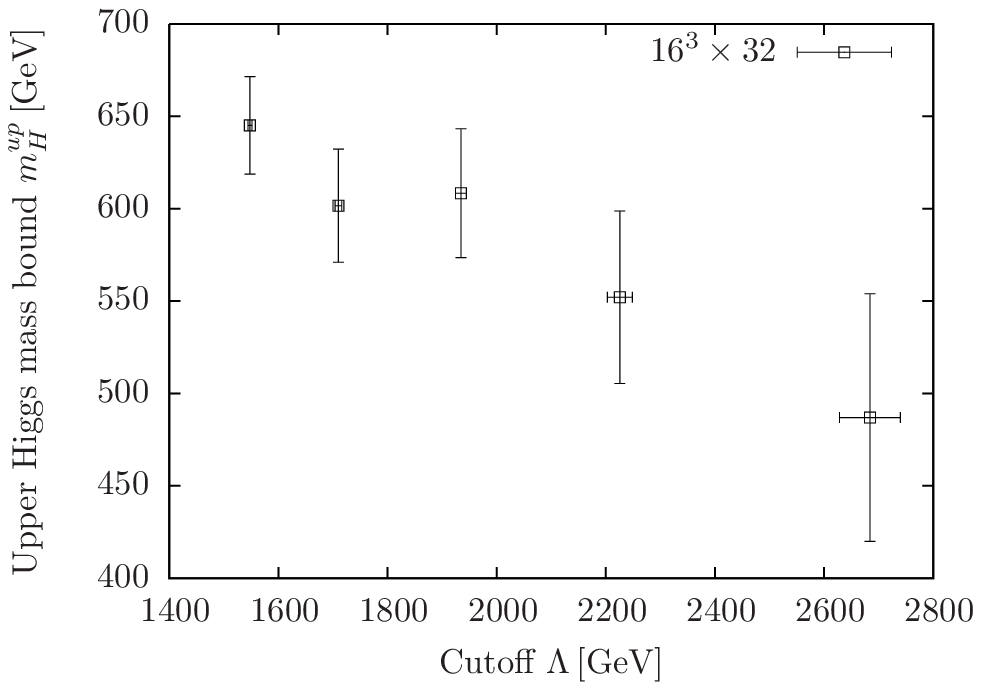}
&
\includegraphics[width=0.48\textwidth]{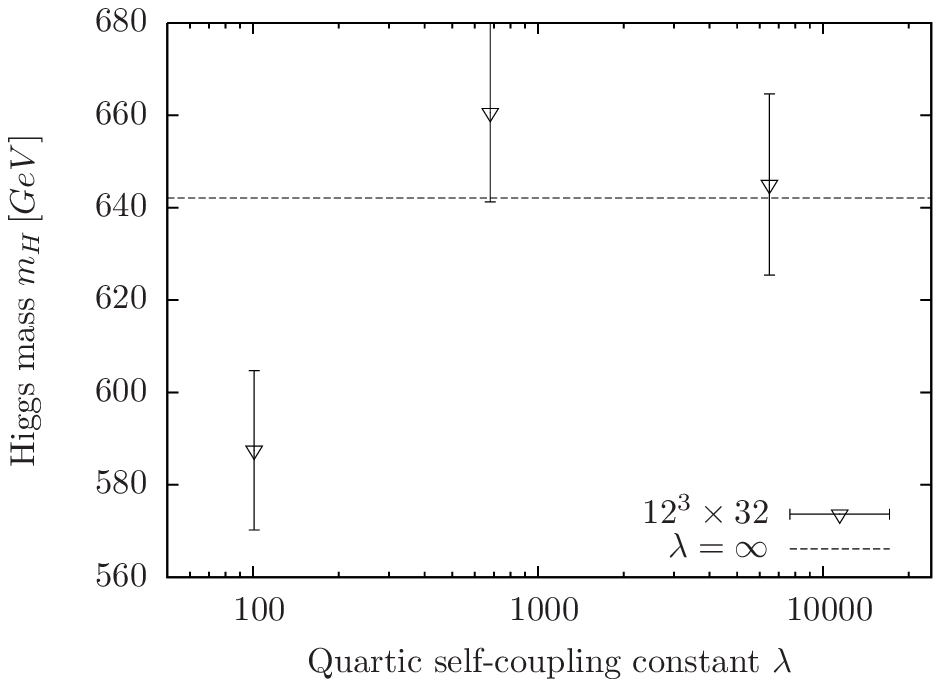}\\
\hs{4mm}(a) & \hs{8mm}(b)  \\
\end{tabular}
\caption{(a) The upper Higgs mass bound $m_H^{up}$ versus the cutoff $\Lambda$ as determined on
a $16^3\times 32$-lattice at infinite quartic self-coupling $\lambda=\infty$.
(b) The dependence of the Higgs mass on the quartic coupling constant $\lambda$ in the strong
quartic coupling regime on a $12^3\times 32$-lattice. The dashed line represents the $\lambda=\infty$
result.}
\label{fig:UpperBound}
\vs{-2mm}
\end{figure}

Finally, we turn towards the determination of the upper Higgs mass bound $m_H^{up}(\Lambda)$. 
First, we check that the largest Higgs masses are indeed obtained at $\lambda=\infty$. This
can be clearly observed in  Fig.~\ref{fig:UpperBound}b where we plot the Higgs mass $m_H$ versus
the quartic self-coupling constant. We therefore derive the upper Higgs mass bounds in the following 
from simulations with infinite self-coupling. In Fig.~\ref{fig:UpperBound}a we present the corresponding
results for the cutoff-dependence of $m_H^{up}(\Lambda)$. As expected the obtained upper mass bounds
fall quickly with increasing cutoff $\Lambda$. Note, however, that the presented results are only 
preliminary, since the considered volumes are rather small and no finite volume effects have been 
studied in this scenario so far.

\bc
\ec

\section*{Acknowledgments}
We thank the "Deutsche Telekom Stiftung" for supporting this study by providing a Ph.D. scholarship for
P.G. We further acknowledge the support of the DFG through the DFG-project {\it Mu932/4-1}.
The numerical computations have been performed on the {\it HP XC4000 System}
at the {\it Scientific Supercomputing Center Karlsruhe} and on the
{\it SGI system HLRN-II} at the {\it HLRN Supercomputing Service Berlin-Hannover}.

\nocite{*}
\bibliographystyle{unsrtOwnNoTitles}  
\bibliography{Proceedings}

\begin{thebibliography}{10}

\bibitem{Holland:2003jr}
K.~Holland and J.~Kuti.
\newblock {\em Nucl. Phys. Proc. Suppl.}, 129:765--767, 2004.

\bibitem{Holland:2004sd}
K.~Holland.
\newblock {\em Nucl. Phys. Proc. Suppl.}, 140:155--161, 2005.

\bibitem{Smit:1989tz}
J.~Smit.
\newblock {\em Nucl. Phys. Proc. Suppl.}, 17:3--16, 1990.

\bibitem{Shigemitsu:1991tc}
J.~Shigemitsu.
\newblock {\em Nucl. Phys. Proc. Suppl.}, 20:515--527, 1991.

\bibitem{Golterman:1990nx}
M.~F.~L. Golterman.
\newblock {\em Nucl. Phys. Proc. Suppl.}, 20:528--541, 1991.

\bibitem{book:Jersak}
A.~K. De and J.~Jers{\'a}k.
\newblock {HLRZ} {J\"u}lich, {HLRZ} 91-83, preprint edition, 1991.

\bibitem{Luscher:1998pq}
M.~L{\"u}scher.
\newblock {\em Phys. Lett.}, B428:342--345, 1998.

\bibitem{Ginsparg:1981bj}
P.~H. Ginsparg and K.~G. {Wilson}.
\newblock {\em Phys. Rev.}, D25:2649, 1982.

\bibitem{Gerhold:2007yb}
P.~Gerhold and K.~Jansen.
\newblock {\em JHEP}, 09:041, 2007.

\bibitem{Gerhold:2007gx}
P.~Gerhold and K.~Jansen.
\newblock {\em JHEP}, 10:001, 2007.

\bibitem{Fodor:2007fn}
Z.~Fodor, K.~Holland, J.~Kuti, D.~Nogradi, and C.~Schroeder.
\newblock {\em PoS}, LAT2007:056, 2007.

\bibitem{Gerhold:2007pj}
P.~Gerhold and K.~Jansen.
\newblock {\em PoS}, LAT2007:075, 2007.

\bibitem{Neuberger:1998wv}
H.~Neuberger.
\newblock {\em Phys. Lett.}, B427:353--355, 1998.

\bibitem{Veltman:1980mj}
M.~J.~G. Veltman.
\newblock {\em Acta Phys. Polon.}, B12:437, 1981.

\end{thebibliography}

\end{document}